\documentclass[fleqn,10pt]{wlscirep}
\usepackage[utf8]{inputenc}
\usepackage[T1]{fontenc}
\usepackage{multirow}
\usepackage{adjustbox}
\title{Monitoring behavioural responses during pandemic via reconstructed contact matrices from online and representative surveys}

\author[1,2,3,+]{J\'ulia Koltai}
\author[3,4,+]{Orsolya V\'as\'arhelyi}
\author[5]{Gergely R\"ost}
\author[3,6,+,*]{M\'arton Karsai}

\affil[1]{\small Computational Social Science – Research Center for Educational and Network Studies, Centre for Social Sciences, Budapest, H-1097, Hungary}
\affil[2]{\small Faculty of Social Sciences, E\"otv\"os Lor\'and University, Budapest, H-1117, Hungary}
\affil[3]{\small Department of Network and Data Science, Central European University, Vienna, A-1100, Austria}
\affil[4]{\small Center for Interdisciplinary Methodologies, University of Warwick, Coventry, United Kingdom}
\affil[5]{\small Bolyai Institute, University of Szeged, Szeged, H-6720, Hungary}
\affil[6]{\small Alfr\'ed R\'enyi Institute of Mathematics, Budapest, H-1053, Hungary}
\affil[+]{these authors contributed equally to this work}
\affil[*]{\small Corresponding author: karsaim@ceu.edu}


\begin{abstract}
The unprecedented behavioural responses of societies have been evidently shaping the COVID-19 pandemic, yet it is a significant challenge to accurately monitor the continuously changing social mixing patterns in real-time. Contact matrices, usually stratified by age, summarise interaction motifs efficiently, but their collection relies on conventional representative survey techniques, which are expensive and slow to obtain. Here we report a data collection effort involving over $2.3\%$ of the Hungarian population to simultaneously record contact matrices through a longitudinal online and sequence of representative phone surveys. To correct non-representative biases characterising the online data, by using census data and the representative samples we develop a reconstruction method to provide a scalable, cheap, and flexible way to dynamically obtain closer-to-representative contact matrices. Our results demonstrate the potential of combined online-offline data collections to understand the changing behavioural responses determining the future evolution of the outbreak, and inform epidemic models with crucial data.
\end{abstract}

\begin{document}

\flushbottom
\maketitle
\thispagestyle{empty}

The spread of directly transmitted diseases such as COVID-19 is largely driven by social interactions and mixing patterns of people~\cite{mossong2008social,rea2007duration,brankston2007transmission}. While person-to-person transmission typically occurs in close contacts,~\cite{musher2003contagious,tellier2006review}, local transportation, commuting, or global travels allow the disease to reach distant territories. Mobility patterns of entire populations can be traced from data coming from transportation or personal digital devices~\cite{vespignani2009predicting}, yet the observation of social interactions is still not obvious. The estimation of interactions and mixing patterns via social proximity, commonly coded as contact matrices~\cite{mossong2008social,fumanelli2012inferring,prem2017projecting}, is difficult, especially when we can only observe a fraction of the population. Even the recently developed contact tracing apps~\cite{ferretti2020quantifying,salathe2020covid,allen2020population} may fail this challenge as they collect too sparse data due to low adoption rate~\cite{wiertz2020predicted,mclachlan2020fundamental}, while they may not assure to keep people's privacy data intact~\cite{bengio2020inherent}. By combining anonymous online data collection techniques with conventional, representative sample based survey methods, we propose a privacy protecting, dynamic, economical, scalable and efficient solution to this problem. Our newly developed large-scale online data collection method, similarly to any other method based on voluntary participation, suffers from unrepresentativity. To overcome this limitation, we developed a detailed weighting methodology using the large-scale online data, and a smaller-scale representative sample simultaneously. This methodology solves the puzzle how voluntary online questionnaires may produce more valid and dynamic contact matrices to inform epidemic models.

The simplest approach to model an epidemic assumes that  contacts between any two individuals occur randomly with equal probability. This so called \emph{homogeneous mixing} assumption dominated the early years of mathematical and computational epidemiology and lead to the seminal results on the dynamics of infectious diseases~\cite{hethcote2000mathematics}. However, %soon it has been realised that 
the heterogeneity of populations called for more refined assumptions to bring the models closer to reality. One successful direction assumes \emph{networked populations} where the social interaction structure of people is taken as the underlying skeleton for epidemic transmission~\cite{pastor2015epidemic}. Social networks commonly appear with various structural heterogeneity~\cite{vega2007complex}, which crucially amplify the chances of global spreading scenarios~\cite{pastor2015epidemic} while making them easier to immunise~\cite{wang2016statistical} in case their global structure is known. However, collecting data about the precise social network of a large population is difficult. Thus, a middle way approach between homogeneously mixed and networked populations is necessary, which is proposed by \emph{contact matrices} representing the aggregated probabilities that different groups of people are in contact with each other~\cite{mossong2008social,fumanelli2012inferring,prem2017projecting}. Most commonly, contacts between age groups are considered, but family structure, gender, education, and other socio-demographic variables have also been used for such stratification~\cite{melegaro2011types,iannelli2005gender,beraud2015french}. The advantages of contact matrices are manifold, as they can be easily integrated to conventional mathematical frameworks to describe the dynamics of an epidemic. Further, they are privacy preserving as they only record aggregated information, yet effectively breaking the homogeneous mixing assumption within a population. They can be dynamically collected and re-scaled to simulate the effects of social distancing or the isolation of different groups for scenario testing of epidemic outcomes.

International and national efforts were implemented worldwide to estimate locally relevant contact matrices for epidemic modelling. One of the largest and earliest effort was carried out by Mossong \emph{et al.} in the POLYMOD project~\cite{mossong2008social}, where in eight European countries 7290 participants were asked to provide their daily contact data to estimate the aggregated age contact matrices. Following these efforts similar studies~\cite{hoang2019systematic} have been conducted in various other countries around the world~\cite{klepac2020contacts,jarvis2020quantifying,beraud2015french,read2014social,zhang2020changes,fu2012representative,leung2017social,ibuka2016social,horby2011social,de2018characteristics,melegaro2017social,kiti2014quantifying,ajelli2017estimating,grijalva2015household}, while several contact matrix estimation methods were also developed~\cite{fumanelli2012inferring,arregui2018projecting}. One important study was published by Prem \emph{et al.}~\cite{prem2017projecting}, who, based on the POLYMOD results and local census data, estimated the contact matrices of 152 countries by using Markov Chain Monte Carlo simulation. All these studies were established on a few paradigms of data collection methods~\cite{read2012close,mccaw2010comparison,beutels2006social}. Several questionnaire based data collection campaigns were carried out using CATI, CAWI or CAPI survey methodologies~\cite{beraud2015french,kiti2014quantifying,ajelli2017estimating}. They commonly collected easily interpretable data, sometimes from representative samples using careful sampling design. Nevertheless, all of them suffered from limited sample size, high cost of data collection, and, except some recent examples~\cite{zhang2020changes}, as they were cross-sectional studies, they completely missed to capture any dynamical change of contact patterns during normal or pandemic periods. On the other hand, online questionnaires and behavioural data collection apps may open new ways to solve these problems. They can reach large populations up to millions of people, while collecting data dynamically, even with changing content, for relatively small costs. However, they may press on privacy issues and due to the voluntary participation, they fall short on providing a representative sample of the observed population. The later crucially limits their direct applicability; as any interpretation drawn from their results need to be handled with caution. Thus, the question remains, how can one exploit all the advantages what online data collection methods provide, while ensuring the privacy of the respondents and the representativeness of the data collected?

\subsubsection*{Actual circumstances}

The recent COVID-19 pandemic called for an immediate answer to this question. In the early days of March 2020, as the COVID-19 pandemic started to unfold in Hungary, a collective action of scientists has been called for the development of country specific epidemic models. This effort was supported by a never seen data sharing initiative by mobile phone providers and health authorities to help realistic data-driven modelling approaches. However, one important data was missing from the beginning: the spatially and demographically detailed age mixing patterns of the country's population. Although estimated~\cite{prem2017projecting} contact matrices were available for Hungary from earlier periods, the actual challenge was to monitor the changes in contact patterns and to measure the societal responses like social distancing or self-protection to nationwide regulations. The Hungarian Data Provider Questionnaire ("Magyar Adatszolg\'altat\'o K\'erd\H{o}\'iv" - MASZK)~\cite{maszk} was developed for these purposes as a voluntary and anonymous online survey, designed by scientists and software engineers~\cite{maszk:team}, as part of a larger project aiming the observation and modelling of the unfolding COVID-19 pandemic in Hungary~\cite{rost2020early}. Beyond collecting static information about the respondents' demography, domicile, education level, or family structure, the primary goal of the questionnaire was to monitor the daily changes in the contact pattern of people in order to calculate the age contact matrices in real time. Additionally, dynamic data was collected about the respondents' employment status, working conditions, physical and mental well-being, and their compliance with recommended self-protection measures during the months of emergency state and beyond. This rolling anonymous online data collection campaign is ongoing up to date (Spring 2021) and reached over $2.3\%$ of the population in Hungary recording over $405,000$ questionnaires from more than $226,000$ individuals, mounting up to the largest data ever collected for this purpose, to our knowledge.

\subsubsection*{Problem and focus}

However, as participation was voluntary, just as any data collected in similar ways, the obtained dataset was not representative for the population of Hungary. To estimate the level and dimensions of unrepresentativity, we performed parallel data collection campaigns based on the same questionnaire, but conducted on a representative sample of $1,500$ people with CATI (computer assisted telephone interviewing) survey methodology. More precisely, additionally to the online survey, we conducted a cross-sectional representative survey in each month from the beginning of the pandemic, in which we measured the actual and pre-COVID-19 contact patterns of participants, and collected all other information recorded in the current online questionnaire. Through the combined analysis of the online and offline data, we evaluated the results of the large online survey and identified its most severe non-representative biases. To account for these biases, we developed a pipeline using iterative proportional fitting\cite{bishop2007discrete} to weight the non-representative data in order to provide more representative contact matrices. This method supports the more realistic measurement of age contact matrices of a whole population while keeping the advantages (like cost-efficiency, scalability and detailed dynamics) of the online data collection. To describe our results, first we briefly summarize the structure and the content of the questionnaire and explain our data collection methods in details. Subsequently we introduce our methodology about the weighting of age contact matrices collected online, in which the dimension of weights are derived from representative data collections conducted in the same period. Finally, we demonstrate our methodology on contact matrices observed during the first wave of the COVID-19 pandemic in Hungary.

\section*{Results}

\subsection*{Data collection}

\paragraph*{The MASZK questionnaire}

The primary purpose of our questionnaire was to dynamically estimate the age contact matrices of people in different environments (like home, work, school, or elsewhere). For this very reason, we asked the respondent about the number of people from different age groups, with whom they had contacts with. First, we recorded \emph{reference contact patterns} by asking respondents about their contacts during a typical weekday and weekend before the COVID-19 outbreak in Hungary (13th March 2020). Second, we recorded \emph{actual contact patterns} of participants by asking them about their contact activities on the day before their actual response. We classified close contacts as \emph{physical contacts} (direct physical contacts without using personal protective equipment), and \emph{proxy contacts} (two persons stayed closer than 2 meters to each other at least for 15 minutes)~\cite{ecdccontact}. Individual contact patterns were recorded as the approximate number of contacts between the ego and their peers from different age groups of $0-4$, $5-14$, $15-29$, $30-44$, $45-59$, $60-69$, $70-79$, and $80+$. For the sake of potential adoption of our method and reproducibility of results we share the core part of our questionnaire including the essential questions for our analysis in the Supplementary Information (SI)~\cite{MASZKquest}.

\begin{figure}[ht]
\centering
\includegraphics[width=\linewidth]{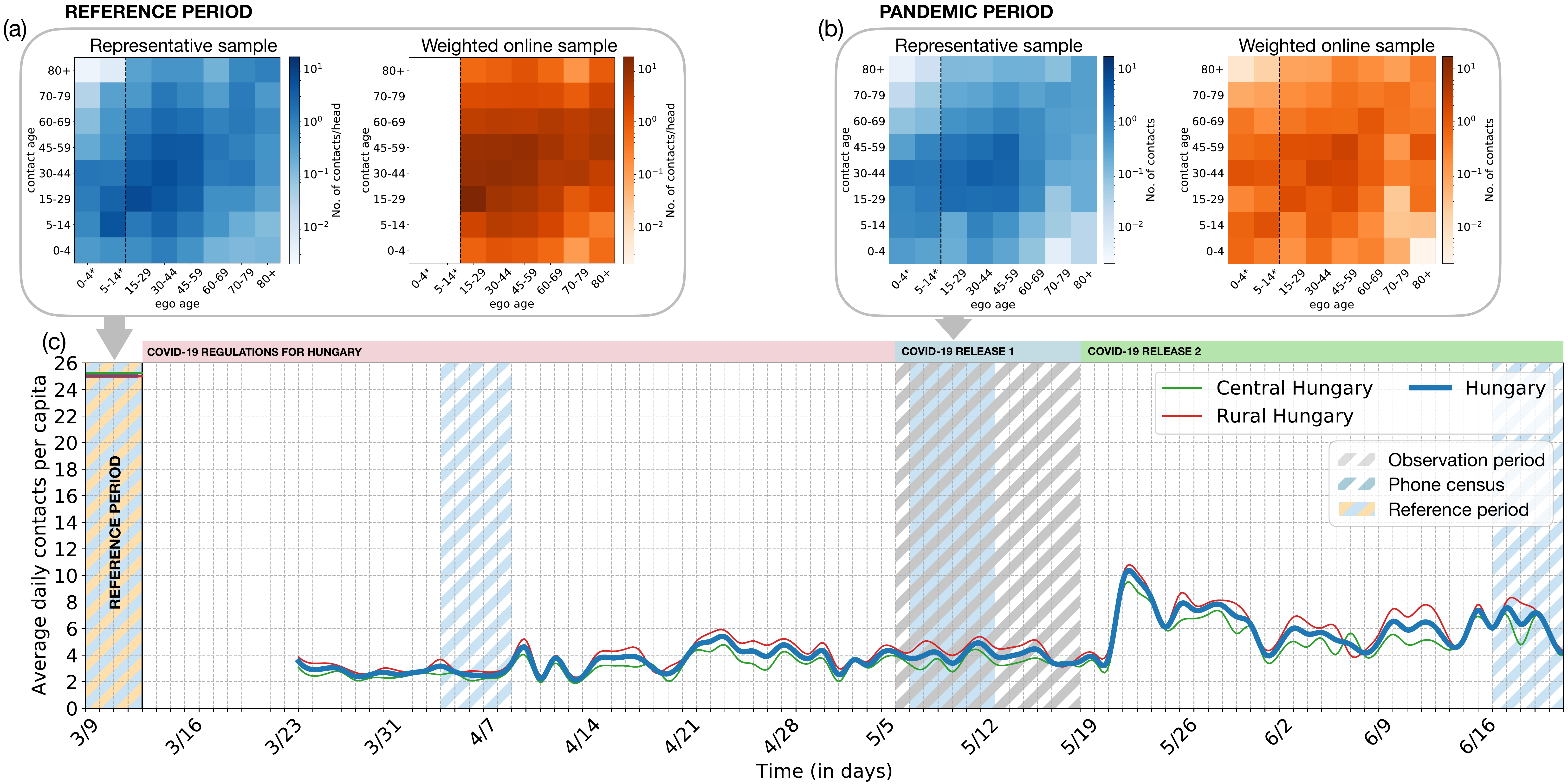}
\caption{\textbf{Contact dynamics, representative and reconstructed age contact matrices.} Age contact matrices measured during the (a) reference and (b) pandemic period via CATI survey methodology on a representative sample (blue) and via weighted non-representative online data collection (orange) after reconstruction (for methodology see section on Construction of age contact matrices). Data for children under $18$ (indicated with asterisk and vertical dashed lines) could not be collected directly due to privacy regulations, thus our data cannot provide a representative sample for the first two age groups. (c) Timeline of early pandemic regulations in Hungary and the average number of per capita daily proxy social contacts in rural areas (solid green line), the central area (red solid line) of Hungary, and in the whole country (blue solid line). While online data collection was continuously ongoing after the 23rd March 2020, representative data via telephone surveys were collected during the periods assigned by diagonal shading. Blue shades indicate telephone census collection, while grey shades cover the online observation period of the actual study. Both methods retrospectively recorded the contact patterns from the reference period (before 13th March 2020), except for age groups under $15$ in the online questionnaire.}
\label{fig:1}
\end{figure}

\paragraph*{Online data collection}

MASZK was originally developed as an online survey\cite{maszk}, and was later published as a mobile phone application~\cite{maszk:app}. Participation was - and still is - voluntary and the data collection was completely anonymous (for further details see the Methods section). The data collection started on the 23rd March 2020 and is still ongoing (as of Spring of 2021). While keeping the core questionnaire (shared in the SI) intact, the additional content has been adjusted to the actually pressing issues of the pandemic, like work and home office conditions, job security, self-protection practices, or intention for vaccination in case of availability. Respondents were asked to fill out the questionnaire as many days, as they can, providing ongoing relevant information about their contacts. Up to date, the questionnaire has been completed in $405,984$ times by $226,086$ respondents, which accounts for $\sim 2.3\%$ of the population of Hungary. The collected data sensitively reflects public awareness and reactions to national regulations as it can be followed in Fig.~\ref{fig:1}c. During the reference period, until the 13th of March 2020 when the first regulations were announced, the average daily number of proxy social contacts of individuals was measured $\sim 25$. This number dropped radically by $88\%$ to a value $\sim 3$ after a national lock-down was introduced. Subsequently, the lock-down was lifted first in rural Hungary (4th May 2020) and later in the more densely populated central region (18th May 2020). This was followed by a modest increase in the number of social contacts to $\sim 8$, which though never reached its reference value until the end of the observed period (20th June 2020). In this work we analyse a period of consecutive three weeks (29th April to 19th May 2020) during the first relaxation of the restrictive measures, as  both types of data collection campaigns were conducted in these days. Using online surveys we recorded $30,770$ responses from $12,208$ people during this three-week period (see Methods, and SI, Table S1).

\paragraph*{Nationally representative telephone survey}

Additionally to the ongoing online data collection, CATI surveys were conducted by a market research company to ask the same questionnaire on a nationally representative sample of people in each month. The sample size was $1,500$, which is $50\%$ larger than the conventional sample size for nationally representative samples in Hungary. Data collection campaigns were conducted in the beginning of the lock-down period (2-7 April 2020), during the first relaxation period (6-12 May 2020), and in each month after May 2020. In the current work, we analyze the data of the second period, where two-third of the data was collected about weekdays, while one third about weekends (for further details see Methods). Our goal with this data collection method was to obtain more realistic and representative data about the contact patterns of the Hungarian population; and to compare similar data coming from different sources to develop tools for reducing biases inherent in the non-representative sample of voluntary online survey.

\subsection*{Construction of age contact matrices}

In order to construct the age contact matrix of social contacts for the whole population, we collected information about the number of proxy and physical contacts of each respondent $x$ during the reference and actual periods in different settings. For a given social connection type, period, and setting, using the age of the respondents we assigned them into one of eight age groups $A$ (as defined in section The MASZK questionnaire), while doing the same for their contacts too. Thus we received an individual contact matrix $\mathrm{M}^x$ coding for each user $x$ the number of contacts they had with others from age groups $i\in A$. Assuming an individual representative weight $w^x$ for each respondent, we computed a weighted average contact matrix $\left(\mathbf{M}\right)_{ij}$, which was column-wise normalised, thus giving us the weighted average number of contacts between a person from age group $j$ with someone from age group $i$. Note that this matrix is not symmetric, and in case of a fully representative sample, weights would be $w^x=1$, simplifying the computation to a simple averaging process (see Methods).

\subsubsection*{Social-demographic biases}

Despite the many advantages of open online surveys, due to voluntary participation they often record a highly non-representative sample of the observed population, which may cause misleading conclusions about the nature of the epidemic process. To identify the most relevant social-demographic dimensions along which the online survey data is biased, we compare the non-representative online data to the corresponding national census.

In most cases, the tests for representativeness of an online survey focus on standard social-demographic characteristics of the observed sample and the population. On the other hand, in our case those characteristics are relevant, which significantly influence the contact patterns of the respondents. To explore these underlying factors, we performed regression analysis on the proxy contacts of respondents in the representative sample recorded in the actual period (for further details see Methods and SI). As the goal of the regression analysis was to detect influencing factors relevant for a later weighting process, the independent variables of these models were not only limited to those asked in the questionnaire, but also by  data available in the census. Although we could identify several significant dimensions, which significantly affect the contact patterns of people, we could not rule out the possibility, that other dimensions, that were not included in the survey or measured by the census, also influence the contact patterns significantly.

\begin{figure}[ht]
\centering
\includegraphics[width=\linewidth]{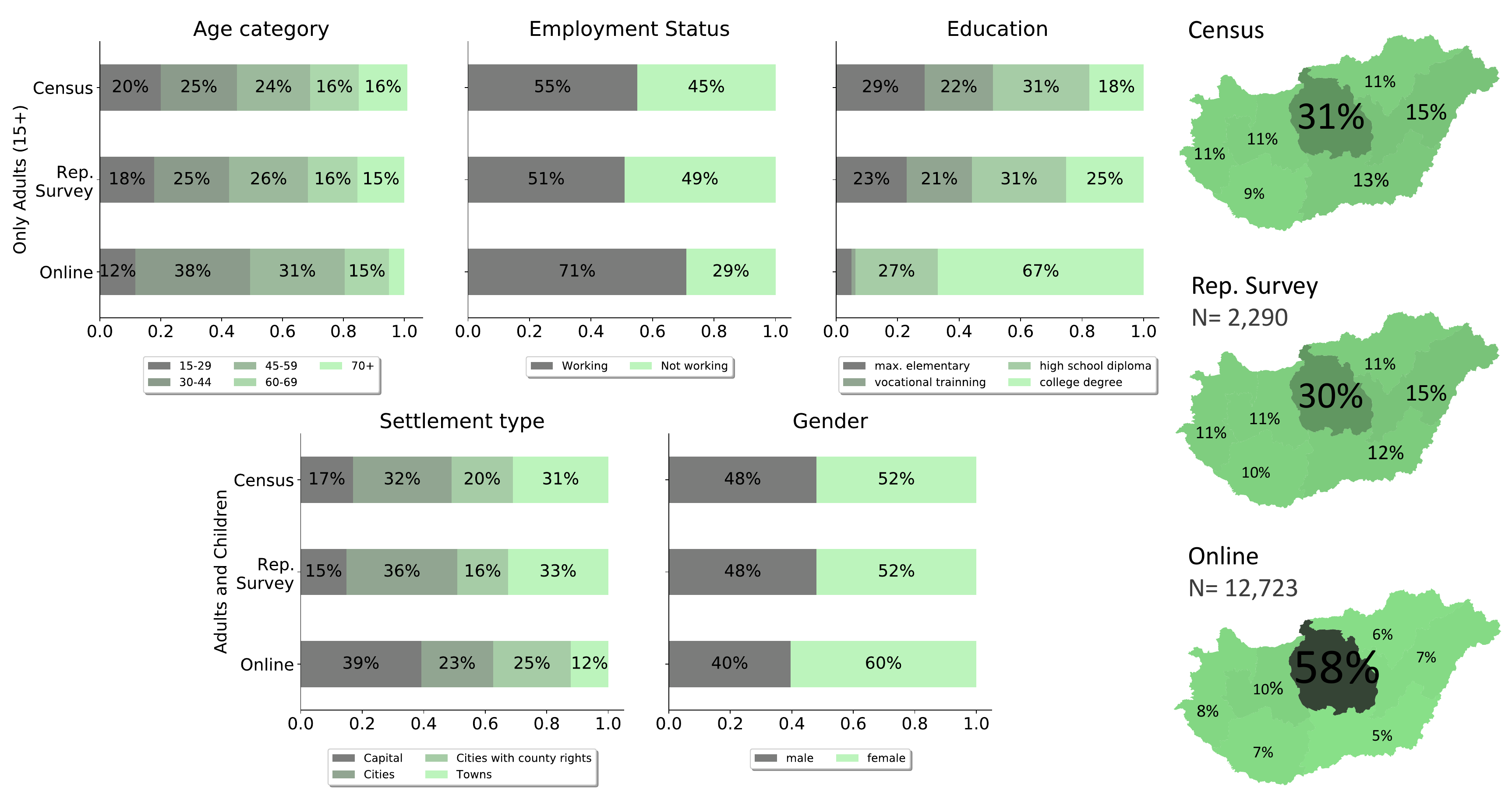}
\caption{\textbf{Descriptive statistics of key demographic variables.} Variable statistics are shown, which were used in the weighting of the raw online data in the representative and in the online survey, compared to the population data of the Hungarian Statistical Office~\cite{KSH}. Note, that statistics showed for age category, employment status and education are based on the adult population of Hungary (15 years old or older), while settlement type, gender and regions covers the entire population. }
\label{fig:2}
\end{figure}

These regression analyses indicated that the age, employment status, education, settlement type, gender and geographical region of the domicile are the most significant social-demographic dimensions along which our online data is non-representative. Indeed, statistics shown in Fig.~\ref{fig:2} evidently demonstrate that while the distributions of the nationally representative phone survey shows very similar values to the population census data provided by the Hungarian Statistical Office~\cite{KSH}, the online survey presents strong biases along these dimensions. Compared to the census data, those who filled out the online survey are more likely to be middle aged, employed, higher educated, live in the capital and more likely to be women. On the other hand, people who are lower educated, older than $70$ years, or live in small settlements like towns are  under-represented. These striking differences demonstrate that the analysis of the raw online survey would lead to biased contact patterns, which are hardly generalizable for the whole Hungarian population.

\subsubsection*{The weighting procedure}

After the detection of those social-demographic variables, which significantly affect the contact patterns observed in the representative survey, we provide a weighting methodology for the online survey to make it more accurate in the measurement of contact patterns of the whole population. The goal of this procedure is to provide an individual weight $w^x$ for every respondent $x$, which indicates how much they are needed to be taken into account in the re-constructed online data to make it representative. Those respondents, who belong to an underrepresented social group get higher weights, while those from over-represented groups get lower ones. From results in Fig,~\ref{fig:2} it is evident, that differences between the online and the census data are quite large. This suggests that individual weights will take values from a very broad range, which is undesirable as extreme weights can result unstable estimations~\cite{lavrakas2008encyclopedia}. Therefore, our weighting methodology needs to meet two goals (1) bringing the online survey data closer to the Hungarian Census, by making it more representative in terms of the identified social-economic dimensions; while (2) keeping the size of the weights in a reasonable range. To meet the second goal, we applied \emph{iterative proportional fitting} (IPF). IPF is a weighting methodology, which adjusts the inner cells of an $n$-dimensional contingency table in a way that it returns the previously provided expected row and column margins \cite{bishop2007discrete}. In our case, the expected margins (the population distributions of the weighting variables) are taken from census data, and the contingency tables, on which we apply the weighting procedure, are derived from the online survey data.

To obtain well fitting weights, which satisfy both of our goals, we built on the age stratified structure of contact matrices. First, as they are built up by age-group-wise normalized vectors for each age group, the relative proportions of age groups can be neglected (not included as expected margins) in the IPF, which considerably decreases the variation of the obtained individual weights. Second, as not necessarily the same dimensions are relevant in each age group (e.g., in some age groups the education level affects contact patterns, in other age groups the geographical location is important.), the identification of relevant weighting dimensions is conducted separately in each age group - which can lead to more realistic weights. The results strengthen this argument as very different social-demographic dimensions affected the total number of actual proxy contacts significantly in different age groups, as summarised in Table~\ref{tab:weigthing} (for margins see SI). 

\begin{table*}
\centering
\begin{tabular}{|c|c|c|c|}
\toprule
Age group & Variables & Weight min. & Weight max. \\
\midrule
0-14 & Gender, Settlement Type, Central / Rural Hungary & 0.29 & 3.23 \\
15-29 & Gender, Central / Rural Hungary & 0.47 & 1.89 \\
30-44 & Region*Work, Region*Settlement   type & 0.44 & 3.35 \\
45-59 & Employment status,   Education*Settlement Type & 0.18 & 6.62 \\
60-69 & Gender, Employment status & 0.50 & 1.73 \\
70+ & Education*Gender, Central / Rural   Hungary, Employment status & 0.04 & 24.79 \\
\bottomrule
\end{tabular}
\caption{\textbf{Table of age groups, the corresponding social-demographic variables and weight limits.} Social-demographic variables are listed for each age group, which were used as margins in the IPF procedure, together with the minimum and maximum values of calculated individual weights. The symbol $*$ indicates interactions between variables. To increase the precision of the weighting procedure, regression analyses targeting the detection of those dimensions, which affect the contact patterns of people, were conducted separately on each age group. The selected dimensions served as expected margins in the IPF procedure. Note that some age groups are merged to make age categories populated enough and to be compatible with the age categories of the census.}
\label{tab:weigthing}
\end{table*}

Compared to standard cell weighting, IPF is less likely to result extremely small or large weights. In our case, after the selection of the relevant dimensions, the IPF process obtained weights, which stayed within the range of $0.04$ and $25.49$ (as presented in Table~\ref{tab:weigthing} with weight distribution summarised in the SI). The closer an individual weight is to one, the more the corresponding individual is representative of their age group - by the listed dimensions. The weight values characterising different age groups can thus disclose, which groups are strongly biased in the online survey as compared to population data. From this perspective of evaluation, the results in Table~\ref{tab:weigthing} suggest that the age groups of 60-69 and 15-29 are the ones closest to the population data of the same age group according to their composition by the listed dimensions. At the same time, the most problematic age group is the 70+, where observed minimum and maximum weights cover the largest range. The larger range of weights can be explained by the self-selection process of respondents, in which older generation is less likely to adopt digital technologies or have internet access, thus, those respondents, who filled out the online questionnaire from this age group are not typical representatives of the whole age group. 

\begin{figure}[ht]
\centering
\includegraphics[width=.8\linewidth]{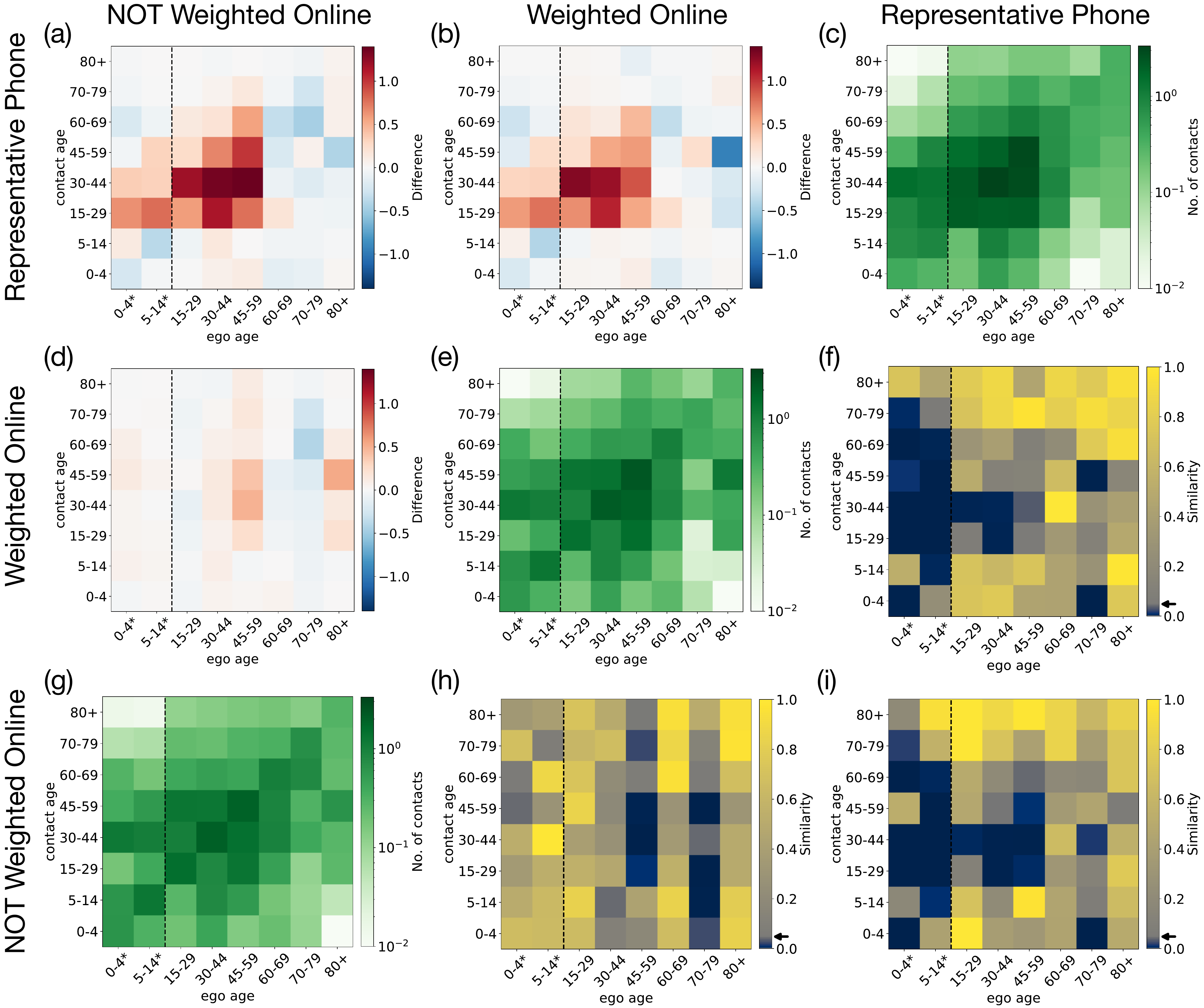}
\caption{\textbf{Results of iterative proportional fitting.} Normalized actual proxy contact matrices (green diagonal), their pairwise difference matrices (above diagonal) and pairwise two-tail T-test results (below diagonal) are depicted for the online non-weighted, online weighted, and representative matrices. In the difference matrices red or blue cells indicate that the source matrix (column label) appeared with higher or lower number of average contact than the target (row label) in a given cell. For results of pairwise two-tail T-tests blue to yellow cells (corresponding to $p>0.05$, assigned by an arrow beside the colorbar) indicate that the given cell is not significantly different in the source (column label) and target (row label) matrices. Data for children under $18$ (indicated with asterisk and vertical dashed lines) could not be collected directly due to privacy regulations, thus our data cannot provide a representative sample for the first two age groups (see Limitations)}
\label{fig:3}
\end{figure}

\subsection*{Reconstructed matrix analysis}

The reconstructed online proxy age contact matrix (panel Fig.~\ref{fig:3}e) appeared with an expected structure very similar to the representative result (panel Fig.~\ref{fig:3}c). It exposes a strong diagonal component induced by age homophily (for annotated matrices see SI), meanwhile it suggests larger contact numbers between people of age 15-59, including the employed population of the country. These matrices were recorded during the period in May 2020, when schools were closed in Hungary. This is reflected in the higher contact numbers between the youngest age groups and their parents' generation from the age group of 30-44. However, if we compare the representative or the reconstructed (weighted) matrices to their corresponding reference period measures (see Fig.~\ref{fig:1}a and b), we evidently see the radical decrease in the number of contacts (darker shades for reference period and lighter for the later one) and the closure of schools significantly reducing the number of homophilic contacts between children of age 5-14 as compared to the reference period.

To quantify the precision of our reconstruction method we compare the raw (not weighted) and reconstructed online proxy contact matrices to the corresponding representative matrix. Although we have demonstrated that the IPF method provides weights within a reasonable range, it is still not evident, which age cells changed the most by the weighting, and which of them became closer to their representative value due to the the reconstruction. In the diagonal of Fig.~\ref{fig:3} we depict the three actual proxy contact matrices built from the representative survey (Fig.~\ref{fig:3}c), from the reconstructed (weighted) online survey (Fig.~\ref{fig:3}e) and the raw (not weighted) online survey (Fig.~\ref{fig:3}g). First, in the upper diagonal, we compare these matrices by calculating their pairwise differences (see Fig.~\ref{fig:3}a, b and d). The difference between the representative survey and the raw online data (Fig.~\ref{fig:3}a) shows that middle-aged respondents of the online data collection had higher number of average contacts with young and middle aged adults than the respondents of the representative survey. Meanwhile, the non-representative online data collection underestimates the number of contacts of elderly people with others of similar age old. However, while the absolute difference in the total number of contacts between the representative and the not weighted online survey was $16.4$, after reconstruction this difference between the representative and weighted online matrices reduced to $14.9$), which corresponds to a $9.13\%$ increase in Relative Accuracy Gain (for precise definition see Methods). Our weighting method performs the best in cases, when the difference between two matrix cells is close to 0 (white in Fig.~\ref{fig:3}b), like in case of the 60-69 years old egos and their 30-44 years old alters. The difference matrix of the non-weighted and weighted matrices depicts the effect of the reconstruction process on the online matrix (see Fig.~\ref{fig:3}d). Although the magnitudes of differences are not large, certain heterogeneities are visible, like the decrease of contact numbers between middle age people and the increase of contacts between 70-79 years old egos and similar others after the reconstruction.

To further quantify the goodness of the weighting in detail, we tested if a cell of a contact matrix is significantly different from the same cell of another contact matrix. Each cell of a contact matrix $\mathbf{M}_{ij}$ appears as the average of the distribution of the number of contacts between the age-group $j$ of a respondent and the age group $i$ of their peers. Thus we can perform a pairwise two-tailed independent sample T-test for each cell to see whether the population means of two groups corresponding to respective cells measured in different contact matrices are significantly different from each other~\cite{david1997paired}. These tests show if the differences presented in the upper diagonal of the figure are statistically significant, or just the results of estimation uncertainties.

In the visualisations of the lower diagonal panels of Fig.~\ref{fig:3}, yellow cells correspond to $p>0.05$ values ($p=0.05$ is indicated by arrows near colorbars) suggesting that average contact numbers between the corresponding age groups are \emph{not} significantly different in the two data sources. For example, this is the case in the cell of egos from age group 45-59, and their peers from 15-29 in Fig.~\ref{fig:3}f, which shows the results of the significance tests comparing the values of the representative and the weighted online matrices. This result suggest that the average contact number between the 45-59, and the 15-29 years old, are not significantly different in the representative and in the weighted online matrices. To check the robustness of our matrix reconstruction method, we performed the same significance test between the raw (not weighted) online matrix and the representative matrices (Fig.~\ref{fig:3}i). Comparing its results to the results of the weighted and representative matrices (Fig.~\ref{fig:3}f), the number of cells, which are not significantly different increased by $6.38\%$ in the latter (from from 44 to 47), while the range of similarity has also elevated (indicated by more yellow cells). Meanwhile, from the T-test results between the raw (not weighted) and weighted online matrices (see Fig.~\ref{fig:3}h) it is evident that the weighting helped to capture the contact patterns better in the reconstructed matrix, especially in case of the active population (30-59) with same-age and older people, and the contacts of the elderly people (70-79) with younger others. Precise estimation of the contact patterns of these age groups are especially important for predicting the potential number of infected cases, which may end up with severe medical conditions in case of the COVID-19 pandemic~\cite{zhang2020changes}. These results show that the reconstruction caused significant changes in the values of 8 cells out of the 64 and that these changes brought the value of the given cell closer to the representative one in most cases (for exact significance values see SI, Figure S2).

\subsubsection*{Limitations and future directions}

It is very important to emphasize that the comparison of the actual proxy contacts in the representative and weighted/not weighted online matrices does not follow the same logic for children in the first two age groups. Due to data protection regulations, the CATI survey is only representative for the adult population of Hungary and not for children, while the online survey could not involve under age children either. Data of children are based on the responses of adult parents estimating the contact patterns of their own children. This estimation is surely biased as, especially for older children, parents may not be fully aware about all daily social contacts of their children. Consequently, we cannot use the representative sample as a 'gold standard' for these age groups, because the population of children recorded in that data is not representative for the children population of the whole country. Correction of this bias would require a separate data collection campaign involving a representative set of children directly, which in turn would raise challenges to meet privacy regulations of under-aged participants and fall beyond the scope of the actual study. Nevertheless, this explains the larger differences between the online and representative matrices in the first two columns in Fig.~\ref{fig:3} off-diagonal panels. If we do not consider these age groups, the Relative Accuracy Gain of the weighting process increases to $11.92\%$ as the absolute difference in the number of contacts between the representative and weighted online survey decreases to $11.36$ which corresponds to an increase of $8.33\% $ in the number of significantly not different cells. To make this bias evident, we separated the non-representative age groups with a vertical dashed line within the matrices, while indicated by asterisks at the labels in each relevant plot.

Another potential limitation may be rooted in the sampling of the observed population. This issue is present at the online data collection, where the number of responses may vary in time. If the size of the online sample is too small, individual weights would diverge and the reconstructed matrices would suffer from large errors. In the present study, this is not an issue, as in the examined period the number of daily responses were stable and relatively high. However in the case of a longitudinal data collection, these parameters can change due to the varying level of public awareness, political influence, or media campaigns.

Finally, not only the number, but also the composition of the respondents may change in time, thus the precision of the actual weights may decrease. To account for this effect in the dynamical reconstruction of contact matrices, one would need to make a representative data collection periodically, and recompute the relevant dimensions and weights for each period. Although we have collected representative samples in each month since April 2020, the demonstration of dynamical re-weighting is the subject of a future investigation (in preparation). There we also plan to apply more experimental weighting procedures, where we will not only include variables available in the census, but also others only available in the representative data. The goal of these weighting experiments is to increase the Relative Accuracy Gain of the procedure.

\section*{Discussion}

Emergency situations, like the actual COVID-19 pandemic, may induce radical changes in the behavioural patterns of people leading to the reduction and re-organisation of their social interactions\cite{van2020using}. Changes may be induced by external influences such as governmental interventions, or change in employment status, but they may strongly depend also on individual decisions induced by self-, and environment-awareness or risk avoiding behaviour. All these influences have convoluted effects on the size and structure of personal interactions leading to different paths of epidemic transmissions in a connected population \cite{block2020social}. Age contact matrices provide a useful way to summarise and follow such changes in the social fabric at different settings and time. Importantly, they can be further used for more realistic modelling of epidemic spreading. Nevertheless, their collection was rather spurious, expensive, and other than some recent studies~\cite{zhang2020changes}, they were collected during 'normal' times, thus they commonly missed to capture changes in contact patterns during emergency periods.

In this study we provide a feasible alternative approach, which combines the advantages of online data collections with the precision provided by representative telephone surveys. We report here, one of the largest data collected to date to estimate age contact matrices in a single country, reaching over $2.3\%$ of the population of Hungary. As the online data provided a non-representative sample of the population, we developed a methodology to reconstruct closer-to representative contact matrices from the online data by using the simultaneously collected representative samples. This data collection method is not only scalable, flexible in terms of content, and relatively cheap, but it also allows for dynamical estimation of contact matrices with high temporal and spatial resolution.

The reproducibility of our results and the possible adoption of our methods in different countries are primary concerns for us. For these reasons, along this study, we share the core questionnaire for further use~\cite{MASZKquest}, together with the raw, reconstructed, and representative matrices and all supporting data calculated for Hungary. Up to date, our data collection method has been implemented already in Mexico~\cite{maszk:mexico} and Cuba. We hope that it will prove useful to collect relevant data for applied epidemiological modelling in other countries too, and at large, will contribute to the global efforts to fight the actual COVID-19 and any future pandemic.

\section*{Materials and Methods}

\subsection*{Data collection}

\subsubsection*{MASZK online data collection} The online data collection started on the 23rd of March 2020 through the website \hyperlink{https://covid.sed.hu}{covid.sed.hu} and later using a mobile phone app~\cite{maszk:app}. The anonymity of participants was ensured by using encrypted browser cookies to store hashed identifiers locally, while transferring only anonymous encrypted data to a central secure server. Encrypted browser cookies were used for the detection of returning respondent filling out the questionnaire on multiple days. The participants did not have to give any information, which could be used for their re-identification. The data collection was fully complying with the actual European and Hungarian privacy data regulations and was approved by the Hungarian National Authority for Data Protection and Freedom of Information~\cite{naih}. The data collection was accompanied with an ongoing marketing campaign, including regular radio and newspaper interviews, ads on social media platforms, and posters on public transportation, to reach the broadest audience possible. Targeted campaigns were also published with help of national organisations to reach parents, university students, or elderly people.

In this study, we analyse data collected between the 29th of April and the 19th of May 2020 and recorded $30,770$ responses from $12,208$ respondents of the online questionnaire. The questionnaire was constructed by two parts in order to minimise the burden and potential churning (sample attrition) of participants:

\paragraph{Static questionnaire:} It was asked only once upon first response (controlled by encrypted browser cookies) about information, which do not change frequently, like the year the respondent was born, gender, domicile, education level, etc. This static part also included questions about the proxy contact patterns of the respondent during the \emph{reference period}, before the official declaration of the pandemic, 13th of March 2020. We recorded reference contact patterns separately for typical weekdays and weekends of the respondents together with their age and gender detailed household structure.
    
\paragraph{Dynamic questionnaire:} It was asked to be completed ideally every day about the activities of the respondent on the previous day. More specifically, we asked the reasons they were outside, the places they visited, the protections they wore, travel mode they used, the changes in their working conditions, etc. We asked questions about their \emph{proxy} and \emph{physical} social contacts outside their home, at work, or elsewhere; and also about those people, with whom they had contacts at home, but who are not part of their household. For those, who mentioned children under 18 years in their household, more questions were asked about the contact patterns of their children at school or elsewhere. We share the full questionnaire including the essential questions for our analysis in the SI.

\subsubsection*{Nationally representative CATI survey}

A smaller scale, but nationwide representative data collection was also conducted between the 6th and 12th of May 2020 using exactly the same questionnaire taken from the online survey. The data collection was implemented by CATI survey methodology using both landline and mobile phone numbers. A multi-step, proportionally stratified, probabilistic sampling procedure was used for sampling. The sample is representative for the Hungarian population aged 18 or older by gender, age, education and domicile. Sampling errors were corrected using iterative proportional post-stratification weights. After data collection, only the anonymised and hashed data was shared with people involved in the project after signing non-disclosure agreements.

\subsection*{Contact matrix construction}
We categorised people into eight age groups, as defined in the main text, thus constructed $8\times 8$ matrices with column indices corresponding to the age group of our respondents and row indices correspond to the age group of their contacts.
In order to compute the population level age contact matrix, we use a formal description. Let $X$ be the set of respondents (ego), and let $Y$ be the set of individuals who are contacts of some $x\in X$. For a specific $x$, let $N_x \subset Y$ be the set of individuals who are contacts of $x$. We assign by $a(x)\in A=\{1,\dots,8\}$ the age group of an individual $x$. Next we define the matrix $M^{x,y}$  for each $x\in X$ and $y \in N_x$ as follows:
$\left(M^{x,y}\right)_{i,j}=1$ if $a(x)=j$ and $a(y)=i$, and zero otherwise. For an ego $x$ we can now compute its individual contact matrix as $\mathrm{M}^x=\sum_{y\in N_x} M^{x,y}$. Finally, we use an individual weight $w^x$ assigned to each ego, coming from the IPF weighting method described in the main text. This weight effectively describes how much an ego and its contacts should be considered in order to receive a contact matrix for a closer-to-representative population. The population level contact matrix is computed by $$\textbf{M}=\sum_{x \in X} w^x \mathrm{M}^x \big/ \sum_{x \in X} w^x.$$

\subsection*{Selection of the weighting dimensions}
The goal of the weighting process was to correct the unrepresentativeness of the online data without getting very large weights which may lead to large errors in the estimations. However, unlike at a general survey, representativeness in our case was not a general term for the Hungarian population, but was related to their contact patterns. To unfold, which variables are the ones that affect the actual proxy contacts the most in the different age groups, we applied linear regression analysis on the representative survey data for each age group separately. The dependent variable of these regressions was the total number of actual proxy contacts; and the independent variables were those ones, which we measured in the questionnaire and which were also available on a population level from census. The following independent variables were matched these two criteria: region (the seven main geographical region of Hungary where the respondent lives), type of settlement of the domicile, gender, highest level of education, and activity (detailed typology of the work type of the respondent - white or blue collar - or the reason they are not employed). We built three models for each age group. In the first model, only the main effects of these variables were included. In the second model we added the two-way interaction terms of all independent variables. Finally, in the third model we included those interaction terms, where neither the region and activity variables were present - as these are categorical data causing too many parameters in the interactions. This step was done to see clearer signals, where the large number of categories of these two variables does not distort the effect of others.
For each age group, we selected the significant variables and the significant interaction terms as weighting dimensions. If a main effect of a variable was significant, and an interaction term, which was built up by the same variable was also significant, we only included the interaction term, because the margins of the interaction also include the margins of those variables, which build that up.
Based on the results of the regression analyses and of the comparison of the online data with the population data, in some cases, we included the aggregated categories (values) of these dimensions in the weighting procedure. For example, in the case of activity, a binary variable was created, where the two categories showed if the respondent worked or did not work. In the case of geographical region, instead of the original seven categories we used two, which showed if the respondent lived in the central region of the country (which includes the capital), or in another region. The reason for these simplifications was that in these variables, the strongest effects on the contact patterns of the people were manifested along these cleavages.

\subsection*{Relative Accuracy Gain}
We define Relative Accuracy Gain (RAG) in our setting to quantify how much we gain in terms of accuracy to approximate the representative contact matrix due to the weighting procedure of the online contact matrix, as compared to the unweighted case. It is defined as the function of the sum of absolute differences in the total number of contacts between the representative (rs) and the weighted online (ow) and the representative and not weighted (onw) online matrices. More formally
\begin{equation} \label{eq3}
RAG = 1-\left(\frac{\sum|\mathbf{M}_{rs}-\mathbf{M}_{ow}|}{\sum|\mathbf{M}_{rs}-\mathbf{M}_{onw}|}\right),
\end{equation}
where $\mathbf{M}_{rs}$ denotes the actual proxy matrix obtained from the nationally representative survey, $\mathbf{M}_{ow}$ is the weighted actual proxy matrix obtained after reconstruction from the online survey, and $\mathbf{M}_{onw}$ is the not weighted actual proxy matrix measured directly from the online survey.

%\bibliography{sample}

\section*{Acknowledgements}
The authors are very thankful for the COVID-19 development team lead by Vilmos Bilicki from the Department of Software Development at the University of Szeged\cite{maszk:team} and for Eszter Bok\'anyi for the data analysis and her constructive comments. This work was done in the framework of the Hungarian National Development, Research, and Innovation (NKFIH) Fund 2020-2.1.1-ED-2020-00003. JK was supported by the Premium Postdoctoral Grant of the Hungarian Academy of Sciences. MK is thankful for the support from the DataRedux (ANR-19-CE46-0008) project funded by ANR and the SoBigData++ (H2020-871042) project. GR was supported by NKFIH FK 124016,  EFOP-3.6.1-16-2016-00008, and TUDFO/47138-1/2019-ITM.

\section*{Author contributions statement}

J.K., M.K and O.V contributed equally to this work, collected data and analysed the results. All authors reviewed the manuscript. 

\section*{Additional information}

\textbf{Competing interests}
The authors declare no competing interests.

\pagebreak

\hspace{-.2in}{\LARGE\textbf{Supplementary Information}}\\ \\
{\Large\textbf{Monitoring behavioural responses during pandemic via reconstructed contact matrices from online and representative surveys}}\\ \\
{\normalsize J\'ulia Koltai, Orsolya V\'as\'arhelyi, Gergely R\"ost and M\'arton Karsai\footnote{corresponding authors: \href{mailto:karsaim@ceu.edu}{karsaim@ceu.edu}}}
\vspace{1cm}

\section{The MASZK Hungarian Data Provider Questionnaire}

The goal of the Hungarian Data Provider Questionnaire (MASZK) questionnaire  was to dynamically estimate the age contact matrices of people in different settings (like home, work, school, or elsewhere). To collect such data we developed a questionnaire to ask about people's demographic characters, domicile, family structure, health conditions, travel patterns, education level, employment situations and many more. More importantly we asked them about the number of people from different age groups, with whom they had contacts. First, we recorded \emph{reference contact patterns} by asking respondents about their contacts during a typical weekday and weekend before the COVID-19 outbreak in Hungary (13th March 2020). Second, we recorded \emph{actual contact patterns} of participants by asking them to indicate all their contact activities happened on the day before their actual response. We defined contacts in two different ways relevant for possible infection transmission.  Interactions between people without any protection were called \emph{physical contacts}, while \emph{proxy contacts} were identified as if two people stayed closer than 2 meters to each other at least for 15 minutes. Individual contact patterns were recorded as the number of contacts between the ego and their peers from different age groups of $0-4$, $5-14$, $15-29$, $30-44$, $45-59$, $60-69$, $70-79$, and $80+$.

Due to privacy regulations, contact patterns of under-age people was not possible directly. Nevertheless, to collect data about children younger than $18$ years old, we asked respondents living in the same household with an under-age children to estimate their number of contacts in different settings.

For the sake of potential adoption of our method and reproducibility of results we share the questionnaire including the essential questions for our analysis in this repository~\cite{MASZKquest}.

\section{Regression analysis to identify relevant factors for weighting}

The overarching goal of the modelling process on the representative survey data was to identify those variables that can be used to weight our non-representative online data, coming from the MASZK questionnaire. Since our goal was to weight our dataset to be more representative for the number of  proxy contacts of respondents, we first ran regression models to identify those factors that significantly affect the daily number of proxy interactions. 

As the contact matrices contain the average number of proxy contacts for each age group separately, we ran general linear models with identity link function separately for each age-group. In this way, we could chose factors, which significantly affect the number of proxy contacts specifically for the given age-group. This method helps to avoid potential weighting variables, which do not influence the contacts for all age segment – and thus it limits the increment of the standard deviation of the estimation.

The dependent variable of the models was the proxy number of contacts; the independent variables were region and activity type as factors, and gender, education and type of settlements as co-variates. The reason for selecting these independent variables was because there are available census data about the distribution of these attributes, which we could use in the weighting procedure later. Moreover, as these dimensions are commonly recorded in any census, it makes possible to easily apply our method in different countries without measuring expensive representative samples. Additionally to the baseline models, we built extended models, to which we included the interaction terms of the independent variables. For each age group we selected those variables or those interactions for the further weighting procedure, which significantly (on a 0.05 level) affected the proxy number of contacts. In the case of the region variable, the results suggested that the main differences are between Central Hungary and other regions, so in the weighting procedure we treated the region variable as a dummy. Similarly, in the case of the independent variable, which measured activity type, the breakpoints were mostly between the active and not active groups, thus we included this variable into the weighting procedure as a dummy one.
The resulting variables are available in Table 1 in the main text, while the full model tables are presented below in Tables S\ref{tab:filtering}, S\ref{tab:census_margins_1} and S\ref{tab:census_margins_2}.

%\textcolor{green}{[FROM MS] for further details on the regression analysis see the Methods section and the SI}

%\textcolor{green}{[FROM MS] Interestingly, after repeating the regression analysis on each age group separately, indeed very different social-demographic dimensions were found to effect significantly the total number of actual proxy contacts in different age groups, as summarised in Table 1 (for margins used for weighting SI)}

\begin{table}
\footnotesize
\begin{tabular}{rrrr}
\toprule
 & \textbf{N} &  &  \\
 \midrule
All respondents & 13,790 &  &  \\
Children* (0-14) & 1,582 &  &  \\
Adults (15+)  & 12,208 &  &  \\
\midrule
\textbf{Age groups - weighting} & \textbf{N} & \textbf{N after weighting} & \textit{Missing} \\
0-14 & 1,582 & 1,527 & \textit{55} \\
15-29 & 1,617 & 1,286 & \textit{} \\
30-44 & 4,423 & 4,152 & \textit{} \\
45-59 & 3,880 & 3,596 & \textit{} \\
60-69 & 1,712 & 1,611 & \textit{} \\
70+ & 576 & 551 & \textit{1,012} \\
\midrule
\textbf{ALL} & \textbf{} & \textbf{12,723} &  \textit{\textbf{1,067}} \\
\bottomrule
\end{tabular}
\caption{Number of responses after filtering users with not avialable data points. The original dataset contains responses from adults and for children based on parents responses. We applied a weighting methodology called iterative proportional fitting on the online survey to make it more accurate of measuring the contact patterns of the whole population. Since this methodology requires each variable used in the weighting to have non zero entry, we had to drop respondents with no data about variables used in the weighting procedure. See Table S2, and S3 for more details about variables used in the weighting procedure. }
\label{tab:filtering}
\end{table}

\begin{table}
\footnotesize
\adjustbox{max width=\linewidth}{%
\begin{tabular}{cccc}
\toprule
\multicolumn{4}{c}{\textbf{0-15}} \\
\midrule
\multicolumn{4}{c}{\textit{Gender}} \\
 & Male & Female &  \\
N & 865,533 & 812,678 &  \\
\% & 51.6\% & 48.4\% &  \\
\midrule
\multicolumn{4}{c}{\textit{Region}} \\
 & Central & Not Central &  \\
N & 519,266 & 1,198,076 &  \\
\% & 30.2\% & 69.8\% &  \\
\midrule
\multicolumn{4}{c}{\textit{Settlement Type}} \\
N &  &  &  \\
\% &  &  &  \\
\toprule
\multicolumn{4}{c}{\textbf{15-29}} \\
\midrule
\multicolumn{4}{c}{\textit{Gender}} \\
 & Male & Female &  \\
N & 865,533 & 812,678 &  \\
\% & 51.6\% & 48.4\% &  \\
\midrule
\multicolumn{4}{c}{\textit{Region}} \\
 & Central & Not Central &  \\
N & 519,266 & 1,198,076 &  \\
\% & 30.2\% & 69.8\% &  \\
\toprule
\multicolumn{4}{c}{\textbf{30-44}} \\
\midrule
\multicolumn{4}{c}{\textit{Center / Rural   Hungary}} \\
 & Central Hungary*Working & Central Hungary * Not Working &  \\
N & 579,409 & 143,751 &  \\
\% & 25.2\% & 6.3\% &  \\
\midrule
 & Not Central Hungary*Working & Not Central Hungary*Not Working &  \\
 & 1,150,796 & 421,478 &  \\
 & 50.1\% & 18.4\% &  \\
\midrule
\multicolumn{4}{c}{\textit{Region*Settlement   Type}} \\
 & Not Central*County city & Not Central*Other city & Not Central*Town \\
N & 460,025 & 532,181 & 562,267 \\
\% & 20\% & 23\% & 24\% \\
 & Budapest & Central Hungary*County city & Central Hungary*Town \\
N & 435,057 & 203,519 & 102,385 \\
\% & 19\% & 9\% & 4\% \\
\bottomrule
\end{tabular}}
\caption{Population census for variables used for applying the weighting methodology called iterative proportional fitting on the online survey to make it more accurate of measuring the contact patterns of the whole population in age groups $0-15$, $15-29$, $30-44$}
\label{tab:census_margins_1}
\end{table}

\begin{table}
\footnotesize
\adjustbox{max width=\linewidth}{%
\begin{tabular}{cccc}
\toprule
\multicolumn{4}{c}{\textbf{45-59}} \\
\midrule
\multicolumn{4}{c}{\textit{Employment   Status}} \\
 & Working & \multicolumn{2}{l}{Not Working} \\
N & 1,617,252 & 317,408 &  \\
\% & 96\% & 19\% &  \\
\midrule
\multicolumn{4}{c}{\textit{Education*Settlement   Type}} \\
 & Capital* max elementary & Capital* high school diploma & Capital*college degree \\
N & 100,013 & 116,443 & 100,773 \\
\% & 6\% & 7\% & 6\% \\
 & County city max elementary & Capital*high school diploma & Capital *college degree \\
N & 172,846 & 139,471 & 97,680 \\
\% & 10\% & 8\% & 6\% \\
 & Other city * max elementary & Other city*high school diploma & Other city*college degree \\
N & 366,678 & 194,829 & 99,679 \\
\% & 21\% & 11\% & 6\% \\
 & Town*max elementary & Town*high school diploma & Town*college degree \\
N & 453,990 & 154,907 & 62,801 \\
\% & 26\% & 9\% & 4\% \\
\toprule
\multicolumn{4}{c}{\textbf{60-69}} \\
\midrule
\multicolumn{4}{c}{\textit{Gender}} \\
 & Male & Female &  \\
N & 579,975 & 732,233 &  \\
\% & 44\% & 56\% &  \\
\midrule
\multicolumn{4}{c}{\textit{Employment Status}} \\
 & Working & Not working &  \\
N & 334,104 & 953,460 &  \\
\% & 26\% & 74\% &  \\
\toprule
\multicolumn{4}{c}{\textbf{70+}} \\
\midrule
\multicolumn{4}{c}{\textit{Gender*Education}} \\
 & Max. elementary*male & Max. elementary*female & High school diploma*male \\
N & 253,655 & 592,334 & 71,699 \\
\% & 15\% & 35\% & 4\% \\
N & High school diploma*male & College degree*male & College degree*female \\
\% & 64,399 & 70,284 & 49,415 \\
 & 4\% & 4\% & 3\% \\
\midrule
\multicolumn{4}{c}{\textit{Region}} \\
 & Central Hungary & \multicolumn{2}{l}{Not Central Hungary} \\
N & 375,500 & 853,082 &  \\
\% & 30.6\% & 69.4\% &  \\
\midrule
\multicolumn{4}{c}{\textit{Employment Status}} \\
 & Working & Not working &  \\
N & 20,708 & 1,133,441 &  \\
\% & 2\% & 98\% & \\
\bottomrule
\end{tabular}}
\caption{Population census of variables used for applying the weighting methodology called iterative proportional fitting on the online survey to make it more accurate of measuring the contact patterns of the whole population in age groups $45-59$, $60-69$, $70+$}
\label{tab:census_margins_2}
\end{table}

\newpage

\section{Weight distribution of Iterative Proportional Fitting}

\begin{figure}[ht]
\centering
\includegraphics[width=.6\linewidth]{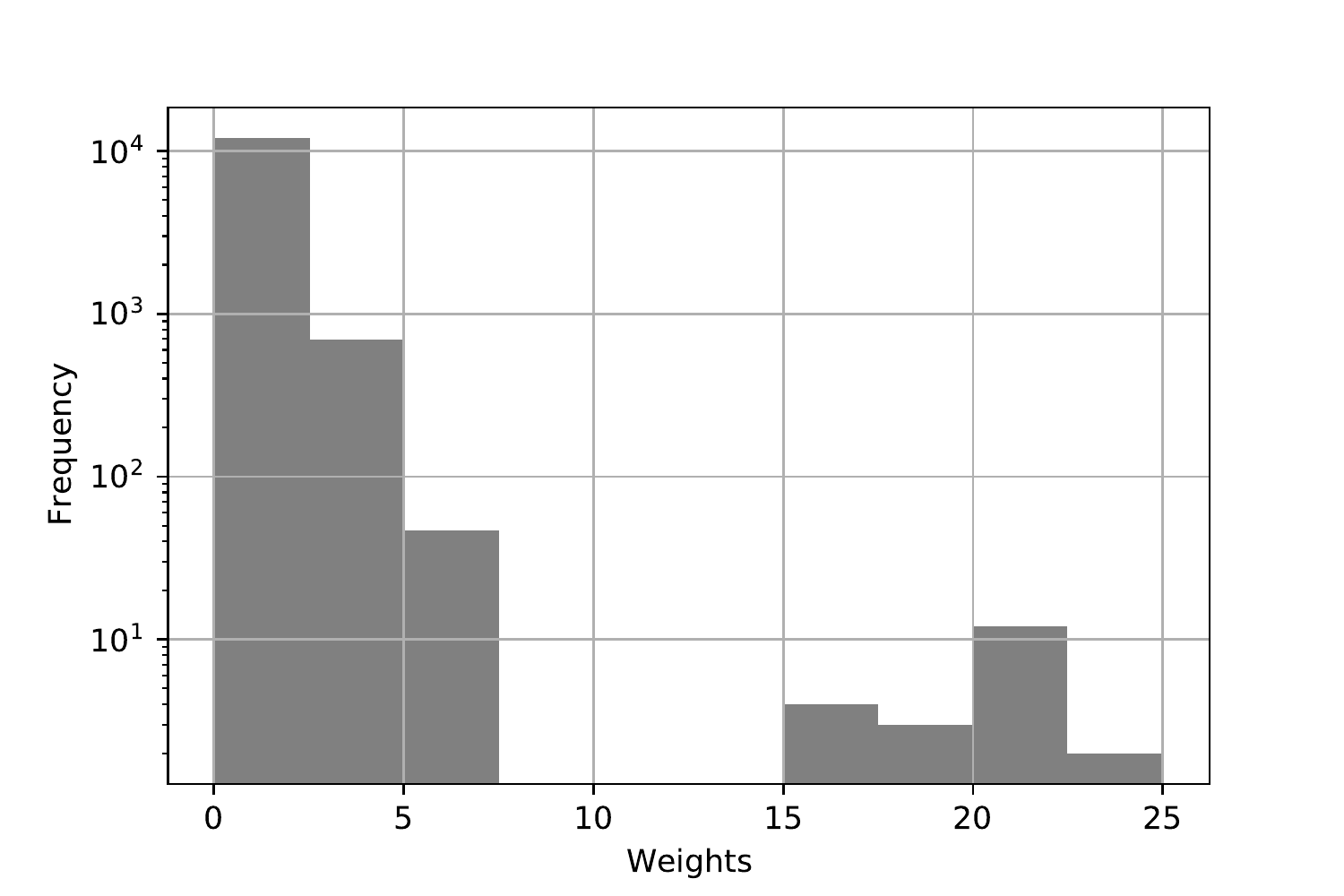}
\caption{Resulting weight distribution of age-stratified iterative proportional fitting (IPF). The obtained weights stayed within the range of 0.04 and 25. 49 (as presented in  Table 1 in the main text}.
\label{fig:SI1}
\end{figure}

Using the detected social-demographic variables effecting significantly the contact patterns, in the MS we described a weighting methodology for the online survey. Our goal was to provide a method which assign a $w^x$ weight to each individual $x$, which are not distributed very broadly, as extreme weights increases the standard errors of the estimates and decrease the accuracy of the estimation. Therefore, our weighting methodology needs to keep the weights in a reasonable range. This was possible by applying \emph{iterative proportional fitting}, which resulted in individual weights distributed over a relatively small range, between $0<w^x<25$ as demonstrated in Fig. S\ref{fig:SI1}).

\section{Annotated matrices}

\begin{figure}
\centering
\includegraphics[width=1\linewidth]{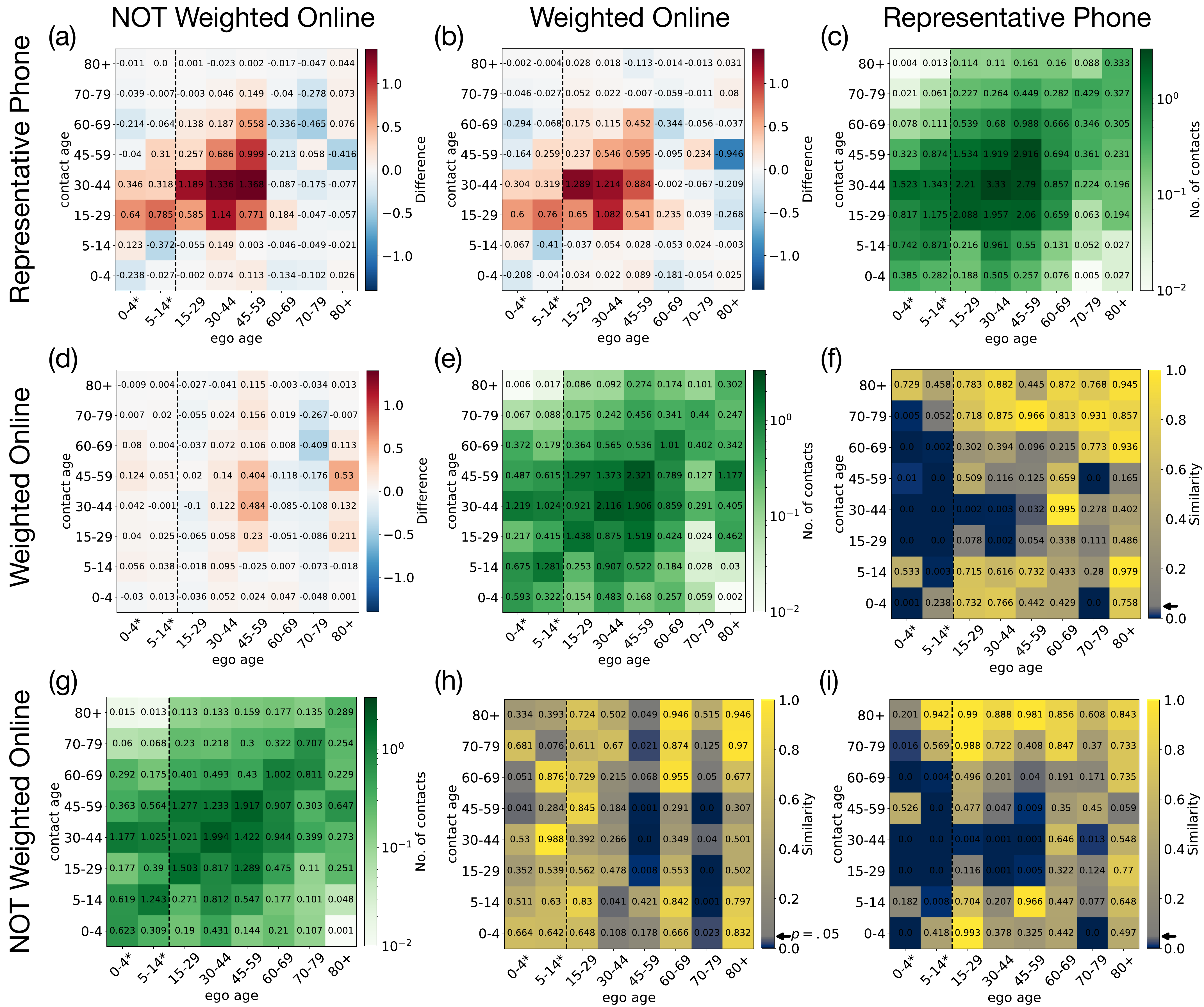}
\caption{\textbf{Annotated matrices.} Normalized actual proxy contact matrices (green diagonal), their pairwise difference matrices (above diagonal) and pairwise two-tail T-test results (below diagonal) are depicted for the online non-weighted, online weighted, and representative matrices. In the difference matrices red or blue cells indicate that the source matrix (column label) appeared with higher or lower number of average contact than the target (row label) at the given cell. For results of pairwise two-tail T-tests yellow to blue cells (corresponding to $p>0.05$, assigned by an arrow beside the colorbar) indicate that the given cell is not significantly different in the source (column label) and target (row label) matrices. Data for children under $18$ (indicated with asterisk and vertical dashed lines) could not be collected directly due to privacy regulations, thus our data cannot provide a representative sample for the first two age groups.}
\label{tab:matrices_annotated}
\end{figure}

To extend our results reported in the main text, here we summarise the measured and reconstructed contact matrices and their comparison in a matrix plot panel, annotated with numerical values. More precisely, in Fig. S\ref{tab:matrices_annotated} we in the diagonal we show the representative, online-weighted, and online-unweighted matrices. Above the diagonal we depict the pairwise differences between these matrices, while below the diagonal we show the pairwise two-tail T-test results.

The raw, reconstructed, and representative matrices are shared as data tables in an online repositories~\cite{MASZKquest}.

\clearpage

\makeatletter
\makeatother
%\bibliography{SI}

\begin{thebibliography}{10}
\urlstyle{rm}
\expandafter\ifx\csname url\endcsname\relax
  \def\url#1{\texttt{#1}}\fi
\expandafter\ifx\csname urlprefix\endcsname\relax\def\urlprefix{URL }\fi
\expandafter\ifx\csname doiprefix\endcsname\relax\def\doiprefix{DOI: }\fi
\providecommand{\bibinfo}[2]{#2}
\providecommand{\eprint}[2][]{\url{#2}}

\bibitem{mossong2008social}
\bibinfo{author}{Mossong, J.} \emph{et~al.}
\newblock \bibinfo{journal}{\bibinfo{title}{Social contacts and mixing patterns
  relevant to the spread of infectious diseases}}.
\newblock {\emph{\JournalTitle{PLoS Medicine}}} \textbf{\bibinfo{volume}{5}}
  (\bibinfo{year}{2008}).

\bibitem{rea2007duration}
\bibinfo{author}{Rea, E.} \emph{et~al.}
\newblock \bibinfo{journal}{\bibinfo{title}{Duration and distance of exposure
  are important predictors of transmission among community contacts of ontario
  sars cases}}.
\newblock {\emph{\JournalTitle{Epidemiology \& Infection}}}
  \textbf{\bibinfo{volume}{135}}, \bibinfo{pages}{914--921}
  (\bibinfo{year}{2007}).

\bibitem{brankston2007transmission}
\bibinfo{author}{Brankston, G.}, \bibinfo{author}{Gitterman, L.},
  \bibinfo{author}{Hirji, Z.}, \bibinfo{author}{Lemieux, C.} \&
  \bibinfo{author}{Gardam, M.}
\newblock \bibinfo{journal}{\bibinfo{title}{Transmission of influenza a in
  human beings}}.
\newblock {\emph{\JournalTitle{The Lancet Infectious Diseases}}}
  \textbf{\bibinfo{volume}{7}}, \bibinfo{pages}{257--265}
  (\bibinfo{year}{2007}).

\bibitem{musher2003contagious}
\bibinfo{author}{Musher, D.~M.}
\newblock \bibinfo{journal}{\bibinfo{title}{How contagious are common
  respiratory tract infections?}}
\newblock {\emph{\JournalTitle{New England Journal of Medicine}}}
  \textbf{\bibinfo{volume}{348}}, \bibinfo{pages}{1256--1266}
  (\bibinfo{year}{2003}).

\bibitem{tellier2006review}
\bibinfo{author}{Tellier, R.}
\newblock \bibinfo{journal}{\bibinfo{title}{Review of aerosol transmission of
  influenza a virus}}.
\newblock {\emph{\JournalTitle{Emerging Infectious Diseases}}}
  \textbf{\bibinfo{volume}{12}}, \bibinfo{pages}{1657} (\bibinfo{year}{2006}).

\bibitem{vespignani2009predicting}
\bibinfo{author}{Vespignani, A.}
\newblock \bibinfo{journal}{\bibinfo{title}{Predicting the behavior of
  techno-social systems}}.
\newblock {\emph{\JournalTitle{Science}}} \textbf{\bibinfo{volume}{325}},
  \bibinfo{pages}{425--428} (\bibinfo{year}{2009}).

\bibitem{fumanelli2012inferring}
\bibinfo{author}{Fumanelli, L.}, \bibinfo{author}{Ajelli, M.},
  \bibinfo{author}{Manfredi, P.}, \bibinfo{author}{Vespignani, A.} \&
  \bibinfo{author}{Merler, S.}
\newblock \bibinfo{journal}{\bibinfo{title}{Inferring the structure of social
  contacts from demographic data in the analysis of infectious diseases
  spread}}.
\newblock {\emph{\JournalTitle{PLoS Computational Biology}}}
  \textbf{\bibinfo{volume}{8}} (\bibinfo{year}{2012}).

\bibitem{prem2017projecting}
\bibinfo{author}{Prem, K.}, \bibinfo{author}{Cook, A.~R.} \&
  \bibinfo{author}{Jit, M.}
\newblock \bibinfo{journal}{\bibinfo{title}{Projecting social contact matrices
  in 152 countries using contact surveys and demographic data}}.
\newblock {\emph{\JournalTitle{PLoS Computational Biology}}}
  \textbf{\bibinfo{volume}{13}}, \bibinfo{pages}{e1005697}
  (\bibinfo{year}{2017}).

\bibitem{ferretti2020quantifying}
\bibinfo{author}{Ferretti, L.} \emph{et~al.}
\newblock \bibinfo{journal}{\bibinfo{title}{Quantifying {SARS-CoV-2}
  transmission suggests epidemic control with digital contact tracing}}.
\newblock {\emph{\JournalTitle{Science}}} \textbf{\bibinfo{volume}{368}}
  (\bibinfo{year}{2020}).

\bibitem{salathe2020covid}
\bibinfo{author}{Salath{\'e}, M.} \emph{et~al.}
\newblock \bibinfo{journal}{\bibinfo{title}{{COVID-19} epidemic in
  {S}witzerland: on the importance of testing, contact tracing and isolation.}}
\newblock {\emph{\JournalTitle{Swiss Medical Weekly}}}
  \textbf{\bibinfo{volume}{150}}, \bibinfo{pages}{w20225}
  (\bibinfo{year}{2020}).

\bibitem{allen2020population}
\bibinfo{author}{Allen, W.~E.} \emph{et~al.}
\newblock \bibinfo{journal}{\bibinfo{title}{Population-scale longitudinal
  mapping of {COVID-19} symptoms, behaviour and testing}}.
\newblock {\emph{\JournalTitle{Nature Human Behaviour}}}
  \textbf{\bibinfo{volume}{4}}, \bibinfo{pages}{972--982}
  (\bibinfo{year}{2020}).

\bibitem{wiertz2020predicted}
\bibinfo{author}{Wiertz, C.}, \bibinfo{author}{Banerjee, A.},
  \bibinfo{author}{Acar, O.~A.} \& \bibinfo{author}{Ghosh, A.}
\newblock \bibinfo{journal}{\bibinfo{title}{Predicted adoption rates of contact
  tracing app configurations-insights from a choice-based conjoint study with a
  representative sample of the {UK} population}}.
\newblock {\emph{\JournalTitle{Available at SSRN 3589199}}}
  (\bibinfo{year}{2020}).

\bibitem{mclachlan2020fundamental}
\bibinfo{author}{McLachlan, S.} \emph{et~al.}
\newblock \bibinfo{journal}{\bibinfo{title}{The fundamental limitations of
  {COVID-19} contact tracing methods and how to resolve them with a {B}ayesian
  network approach}}.
\newblock {\emph{\JournalTitle{Risk Inf. Manage., London, U.K., Tech. Rep.}}}
  \doiprefix\url{10.13140/RG.2.2.27042.66243} (\bibinfo{year}{2020}).

\bibitem{bengio2020inherent}
\bibinfo{author}{Bengio, Y.} \emph{et~al.}
\newblock \bibinfo{journal}{\bibinfo{title}{Inherent privacy limitations of
  decentralized contact tracing apps}}.
\newblock {\emph{\JournalTitle{Journal of the American Medical Informatics
  Association}}}  (\bibinfo{year}{2020}).

\bibitem{hethcote2000mathematics}
\bibinfo{author}{Hethcote, H.~W.}
\newblock \bibinfo{journal}{\bibinfo{title}{The mathematics of infectious
  diseases}}.
\newblock {\emph{\JournalTitle{SIAM Review}}} \textbf{\bibinfo{volume}{42}},
  \bibinfo{pages}{599--653} (\bibinfo{year}{2000}).

\bibitem{pastor2015epidemic}
\bibinfo{author}{Pastor-Satorras, R.}, \bibinfo{author}{Castellano, C.},
  \bibinfo{author}{Van~Mieghem, P.} \& \bibinfo{author}{Vespignani, A.}
\newblock \bibinfo{journal}{\bibinfo{title}{Epidemic processes in complex
  networks}}.
\newblock {\emph{\JournalTitle{Reviews of Modern Physics}}}
  \textbf{\bibinfo{volume}{87}}, \bibinfo{pages}{925} (\bibinfo{year}{2015}).

\bibitem{vega2007complex}
\bibinfo{author}{Vega-Redondo, F.}
\newblock \emph{\bibinfo{title}{Complex social networks}}.
\newblock \bibinfo{number}{44} (\bibinfo{publisher}{Cambridge University
  Press}, \bibinfo{year}{2007}).

\bibitem{wang2016statistical}
\bibinfo{author}{Wang, Z.} \emph{et~al.}
\newblock \bibinfo{journal}{\bibinfo{title}{Statistical physics of
  vaccination}}.
\newblock {\emph{\JournalTitle{Physics Reports}}}
  \textbf{\bibinfo{volume}{664}}, \bibinfo{pages}{1--113}
  (\bibinfo{year}{2016}).

\bibitem{melegaro2011types}
\bibinfo{author}{Melegaro, A.}, \bibinfo{author}{Jit, M.},
  \bibinfo{author}{Gay, N.}, \bibinfo{author}{Zagheni, E.} \&
  \bibinfo{author}{Edmunds, W.~J.}
\newblock \bibinfo{journal}{\bibinfo{title}{What types of contacts are
  important for the spread of infections? using contact survey data to explore
  {E}uropean mixing patterns}}.
\newblock {\emph{\JournalTitle{Epidemics}}} \textbf{\bibinfo{volume}{3}},
  \bibinfo{pages}{143--151} (\bibinfo{year}{2011}).

\bibitem{iannelli2005gender}
\bibinfo{author}{Iannelli, M.}, \bibinfo{author}{Martcheva, M.} \&
  \bibinfo{author}{Milner, F.~A.}
\newblock \emph{\bibinfo{title}{Gender-structured population modeling:
  mathematical methods, numerics, and simulations}} (\bibinfo{publisher}{SIAM},
  \bibinfo{year}{2005}).

\bibitem{beraud2015french}
\bibinfo{author}{B{\'e}raud, G.} \emph{et~al.}
\newblock \bibinfo{journal}{\bibinfo{title}{The {F}rench connection: the first
  large population-based contact survey in {F}rance relevant for the spread of
  infectious diseases}}.
\newblock {\emph{\JournalTitle{PLoS ONE}}} \textbf{\bibinfo{volume}{10}}
  (\bibinfo{year}{2015}).

\bibitem{hoang2019systematic}
\bibinfo{author}{Hoang, T.} \emph{et~al.}
\newblock \bibinfo{journal}{\bibinfo{title}{A systematic review of social
  contact surveys to inform transmission models of close-contact infections}}.
\newblock {\emph{\JournalTitle{Epidemiology (Cambridge, Mass.)}}}
  \textbf{\bibinfo{volume}{30}}, \bibinfo{pages}{723} (\bibinfo{year}{2019}).

\bibitem{klepac2020contacts}
\bibinfo{author}{Klepac, P.} \emph{et~al.}
\newblock \bibinfo{journal}{\bibinfo{title}{Contacts in context: large-scale
  setting-specific social mixing matrices from the {BBC} {P}andemic project}}.
\newblock {\emph{\JournalTitle{medRxiv 2020.02.16.20023754}}}
  (\bibinfo{year}{2020}).

\bibitem{jarvis2020quantifying}
\bibinfo{author}{Jarvis, C.~I.} \emph{et~al.}
\newblock \bibinfo{journal}{\bibinfo{title}{Quantifying the impact of physical
  distance measures on the transmission of {COVID-19} in the {UK}}}.
\newblock {\emph{\JournalTitle{BMC Medicine}}} \textbf{\bibinfo{volume}{18}},
  \bibinfo{pages}{1--10} (\bibinfo{year}{2020}).

\bibitem{read2014social}
\bibinfo{author}{Read, J.~M.} \emph{et~al.}
\newblock \bibinfo{journal}{\bibinfo{title}{Social mixing patterns in rural and
  urban areas of {S}outhern {C}hina}}.
\newblock {\emph{\JournalTitle{Proceedings of the Royal Society B: Biological
  Sciences}}} \textbf{\bibinfo{volume}{281}}, \bibinfo{pages}{20140268}
  (\bibinfo{year}{2014}).

\bibitem{zhang2020changes}
\bibinfo{author}{Zhang, J.} \emph{et~al.}
\newblock \bibinfo{journal}{\bibinfo{title}{Changes in contact patterns shape
  the dynamics of the {COVID-19} outbreak in {C}hina}}.
\newblock {\emph{\JournalTitle{Science}}}  (\bibinfo{year}{2020}).

\bibitem{fu2012representative}
\bibinfo{author}{Fu, Y.-c.}, \bibinfo{author}{Wang, D.-W.} \&
  \bibinfo{author}{Chuang, J.-H.}
\newblock \bibinfo{journal}{\bibinfo{title}{Representative contact diaries for
  modeling the spread of infectious diseases in {T}aiwan}}.
\newblock {\emph{\JournalTitle{PLoS ONE}}} \textbf{\bibinfo{volume}{7}}
  (\bibinfo{year}{2012}).

\bibitem{leung2017social}
\bibinfo{author}{Leung, K.}, \bibinfo{author}{Jit, M.}, \bibinfo{author}{Lau,
  E.~H.} \& \bibinfo{author}{Wu, J.~T.}
\newblock \bibinfo{journal}{\bibinfo{title}{Social contact patterns relevant to
  the spread of respiratory infectious diseases in hong kong}}.
\newblock {\emph{\JournalTitle{Scientific Reports}}}
  \textbf{\bibinfo{volume}{7}}, \bibinfo{pages}{1--12} (\bibinfo{year}{2017}).

\bibitem{ibuka2016social}
\bibinfo{author}{Ibuka, Y.} \emph{et~al.}
\newblock \bibinfo{journal}{\bibinfo{title}{Social contacts, vaccination
  decisions and influenza in {J}apan}}.
\newblock {\emph{\JournalTitle{J Epidemiol Community Health}}}
  \textbf{\bibinfo{volume}{70}}, \bibinfo{pages}{162--167}
  (\bibinfo{year}{2016}).

\bibitem{horby2011social}
\bibinfo{author}{Horby, P.} \emph{et~al.}
\newblock \bibinfo{journal}{\bibinfo{title}{Social contact patterns in vietnam
  and implications for the control of infectious diseases}}.
\newblock {\emph{\JournalTitle{PLoS ONE}}} \textbf{\bibinfo{volume}{6}}
  (\bibinfo{year}{2011}).

\bibitem{de2018characteristics}
\bibinfo{author}{de~Waroux, O. l.~P.} \emph{et~al.}
\newblock \bibinfo{journal}{\bibinfo{title}{Characteristics of human encounters
  and social mixing patterns relevant to infectious diseases spread by close
  contact: a survey in {S}outhwest {U}ganda}}.
\newblock {\emph{\JournalTitle{BMC Infectious Diseases}}}
  \textbf{\bibinfo{volume}{18}}, \bibinfo{pages}{172} (\bibinfo{year}{2018}).

\bibitem{melegaro2017social}
\bibinfo{author}{Melegaro, A.} \emph{et~al.}
\newblock \bibinfo{journal}{\bibinfo{title}{Social contact structures and time
  use patterns in the {M}anicaland {P}rovince of {Z}imbabwe}}.
\newblock {\emph{\JournalTitle{PLoS ONE}}} \textbf{\bibinfo{volume}{12}}
  (\bibinfo{year}{2017}).

\bibitem{kiti2014quantifying}
\bibinfo{author}{Kiti, M.~C.} \emph{et~al.}
\newblock \bibinfo{journal}{\bibinfo{title}{Quantifying age-related rates of
  social contact using diaries in a rural coastal population of {K}enya}}.
\newblock {\emph{\JournalTitle{PLoS ONE}}} \textbf{\bibinfo{volume}{9}}
  (\bibinfo{year}{2014}).

\bibitem{ajelli2017estimating}
\bibinfo{author}{Ajelli, M.} \& \bibinfo{author}{Litvinova, M.}
\newblock \bibinfo{journal}{\bibinfo{title}{Estimating contact patterns
  relevant to the spread of infectious diseases in {R}ussia}}.
\newblock {\emph{\JournalTitle{Journal of Theoretical Biology}}}
  \textbf{\bibinfo{volume}{419}}, \bibinfo{pages}{1--7} (\bibinfo{year}{2017}).

\bibitem{grijalva2015household}
\bibinfo{author}{Grijalva, C.~G.} \emph{et~al.}
\newblock \bibinfo{journal}{\bibinfo{title}{A household-based study of contact
  networks relevant for the spread of infectious diseases in the highlands of
  {P}eru}}.
\newblock {\emph{\JournalTitle{PLoS ONE}}} \textbf{\bibinfo{volume}{10}}
  (\bibinfo{year}{2015}).

\bibitem{arregui2018projecting}
\bibinfo{author}{Arregui, S.}, \bibinfo{author}{Aleta, A.},
  \bibinfo{author}{Sanz, J.} \& \bibinfo{author}{Moreno, Y.}
\newblock \bibinfo{journal}{\bibinfo{title}{Projecting social contact matrices
  to different demographic structures}}.
\newblock {\emph{\JournalTitle{PLoS Computational Biology}}}
  \textbf{\bibinfo{volume}{14}}, \bibinfo{pages}{e1006638}
  (\bibinfo{year}{2018}).

\bibitem{read2012close}
\bibinfo{author}{Read, J.}, \bibinfo{author}{Edmunds, W.},
  \bibinfo{author}{Riley, S.}, \bibinfo{author}{Lessler, J.} \&
  \bibinfo{author}{Cummings, D.}
\newblock \bibinfo{journal}{\bibinfo{title}{Close encounters of the infectious
  kind: methods to measure social mixing behaviour}}.
\newblock {\emph{\JournalTitle{Epidemiology \& Infection}}}
  \textbf{\bibinfo{volume}{140}}, \bibinfo{pages}{2117--2130}
  (\bibinfo{year}{2012}).

\bibitem{mccaw2010comparison}
\bibinfo{author}{McCaw, J.~M.} \emph{et~al.}
\newblock \bibinfo{journal}{\bibinfo{title}{Comparison of three methods for
  ascertainment of contact information relevant to respiratory pathogen
  transmission in encounter networks}}.
\newblock {\emph{\JournalTitle{BMC Infectious Diseases}}}
  \textbf{\bibinfo{volume}{10}}, \bibinfo{pages}{166} (\bibinfo{year}{2010}).

\bibitem{beutels2006social}
\bibinfo{author}{Beutels, P.}, \bibinfo{author}{Shkedy, Z.},
  \bibinfo{author}{Aerts, M.} \& \bibinfo{author}{Van~Damme, P.}
\newblock \bibinfo{journal}{\bibinfo{title}{Social mixing patterns for
  transmission models of close contact infections: exploring self-evaluation
  and diary-based data collection through a web-based interface}}.
\newblock {\emph{\JournalTitle{Epidemiology \& Infection}}}
  \textbf{\bibinfo{volume}{134}}, \bibinfo{pages}{1158--1166}
  (\bibinfo{year}{2006}).

\bibitem{maszk}
\bibinfo{title}{Hungarian data supply questionnaire (maszk) (date of access
  2020.09.28)}.

\bibitem{maszk:team}
\bibinfo{title}{Hungarian data supply questionnaire (maszk) team,
  https://covid.sed.hu/tabs/staff, (date of access 2020.09.28)}.

\bibitem{rost2020early}
\bibinfo{author}{R{\"o}st, G.} \emph{et~al.}
\newblock \bibinfo{journal}{\bibinfo{title}{Early phase of the {COVID-19}
  outbreak in {H}ungary and post-lockdown scenarios}}.
\newblock {\emph{\JournalTitle{Viruses}}} \textbf{\bibinfo{volume}{12}},
  \bibinfo{pages}{708} (\bibinfo{year}{2020}).

\bibitem{bishop2007discrete}
\bibinfo{author}{Bishop, Y.~M.}, \bibinfo{author}{Fienberg, S.~E.} \&
  \bibinfo{author}{Holland, P.~W.}
\newblock \emph{\bibinfo{title}{Discrete multivariate analysis: theory and
  practice}} (\bibinfo{publisher}{Springer Science \& Business Media},
  \bibinfo{year}{2007}).

\bibitem{ecdccontact}
\bibinfo{title}{Surveillance definitions for {COVID-19}, {E}uropean {C}entre
  for {D}isease {P}revention and {C}ontrol,
  https://www.ecdc.europa.eu/en/covid-19/surveillance/surveillance-definitions,
  (date of access 2020.09.28)}.

\bibitem{MASZKquest}
\bibinfo{title}{{MASZK - Hungarian Data Provider Questionnaire},
  \url{https://figshare.com/articles/online_resource/Hungarian_Data_Provider_Questionnaire/13550057}}.

\bibitem{maszk:app}
\bibinfo{author}{Dr. Vilmos Bilicki MASZK Development~Team, D. o. S.~D.,
  University of~Szeged}.
\newblock \bibinfo{title}{Maszk app for android,
  https://play.google.com/store/apps/ (date of access 2020.10.02)}.

\bibitem{KSH}
\bibinfo{title}{Magyar n\'epsz\'aml\'al\'as 2011,
  http://www.ksh.hu/nepszamlalas/ (date of access 2020.12.)}.

\bibitem{lavrakas2008encyclopedia}
\bibinfo{author}{Lavrakas, P.~J.}
\newblock \emph{\bibinfo{title}{Encyclopedia of survey research methods}}
  (\bibinfo{publisher}{Sage Publications}, \bibinfo{year}{2008}).

\bibitem{david1997paired}
\bibinfo{author}{David, H.~A.} \& \bibinfo{author}{Gunnink, J.~L.}
\newblock \bibinfo{journal}{\bibinfo{title}{The paired t test under artificial
  pairing}}.
\newblock {\emph{\JournalTitle{The American Statistician}}}
  \textbf{\bibinfo{volume}{51}}, \bibinfo{pages}{9--12} (\bibinfo{year}{1997}).

\bibitem{van2020using}
\bibinfo{author}{Van~Bavel, J.~J.} \emph{et~al.}
\newblock \bibinfo{journal}{\bibinfo{title}{Using social and behavioural
  science to support {COVID-19} pandemic response}}.
\newblock {\emph{\JournalTitle{Nature Human Behaviour}}} \bibinfo{pages}{1--12}
  (\bibinfo{year}{2020}).

\bibitem{block2020social}
\bibinfo{author}{Block, P.} \emph{et~al.}
\newblock \bibinfo{journal}{\bibinfo{title}{Social network-based distancing
  strategies to flatten the {COVID-19} curve in a post-lockdown world}}.
\newblock {\emph{\JournalTitle{Nature Human Behaviour}}} \bibinfo{pages}{1--9}
  (\bibinfo{year}{2020}).

\bibitem{maszk:mexico}
\bibinfo{title}{{COVID-19} {UNAM}, https://coronavirusapoyamexico.c3.unam.mx/
  (date of access 2020.12.)}.

\bibitem{naih}
\bibinfo{title}{Nemzeti adatv\'edelmi \'es inform\'aci\'oszabads\'ag
  hat\'os\'ag, https://www.naih.hu (date of access 2020.12.)}.

\end{thebibliography}

\end{document}